\documentclass[%
preprintnumbers,
 amsmath,amssymb,
 aps,
]{revtex4-2}

  \def\rtx@apsprfluids{%
 \class@info{APS journal PRFluids selected}
 }%

\usepackage{graphicx}
\usepackage{dcolumn}
\usepackage{bm}
\usepackage{hyperref}
\usepackage[english]{babel}

\begin{document}

\preprint{APS/123-QED}

\title{Interaction of Vertical Convection with an Electromagnetically Forced Flow}

\author{Evgeniy Shvydkiy}
 \email{e.l.shvydky@urfu.ru}
\author{Ivan Smolyanov}%
\affiliation{Department of Electrical Engineering, Ural Federal University}


\author{Egbert Baake}
\affiliation{Institute of Electrotechnology, Leibniz University Hannover}%

\date{\today}

\begin{abstract}



Heat and momentum transfer of low-Prandtl-number fluid ($Pr=0.029$) in a closed rectangular cavity ($100\times60\times10$ mm$^3$) heated at one side and cooled at the opposite side are analyzed. 
The electromagnetic forces into the liquid metal are generated by the travelling magnetic field inductor and directed towards buoyancy forces. 
Large eddy simulations are performed with the Grashof number $Gr$ from $1.9\cdot 10^5$ to $7.6\cdot 10^7$ and the electromagnetic forcing parameter $F$ from $2.6\cdot 10^4$ to $2.6\cdot10^6$. 
An experimental validation of the simulation results of vertical convection and electromagnetically driven flow using GaInSn alloy has been performed.
Three types of flow patterns are obtained for different interaction parameters $ N = F / Gr $: counterclockwise flow, clockwise flow, and coexistence of two vortices.
Analysis of the Reynolds number shows that the transition zone from natural convection to electromagnetic stirring lies in the range $0.02<F/Gr<0.07$ and two braking modes are found.
The transition point between the convective heat transfer regimes is found for $ F / Gr $ around 1.
The analysis of isotherms deformation showed that in such convective systems it is possible to achieve minimum deviation of the isotherm shape from a straight line in the range of $ 0.05 <F/Gr <0.2 $.


\end{abstract}

\maketitle


\section{\label{intro}Introduction}


Natural thermal convection exists in many environmental or industry processes.
For research on various issues in convection, \textit{Rayleigh-Bénard convection} (RBC) is the most fundamental and classical configuration \cite{YangPhysRevFluids2021}.
In RBC, bottom liquid layers are heated up and bottom ones are cooled to produce planar vertical temperature gradient for motion of the liquid \cite{RevModPhys.81.503,HePhysRevLett2012,ChillEPJE2012}.
If the Rayleigh-Bénard convection is rotated at 90 degrees, then a horizontally applied temperature gradient becomes  vertical, therefore, this leads to the occurrence of \textit{vertical convection} (VC).
Since thermally driven VC is the flow in a cavity heated on one side and cooled on the other side \cite{wangJFM2021}. Thermal convection inevitably arises in the case of a horizontal temperature gradient.
It is also worth noting that the VC and RBC are the two limiting cases in the \textit{inclined convection} \cite{FrickEPL2015,shishkina_horn_2016,TeimurazovPRF2017,ZwirnerJFM2020}.


One of the first studies of vertical convection was carried out by G.K. Batchelor \cite{Batchelor1954}. The main interest of his work was in studying heat transport through double layer windows.
Later, a number of works on the VC topic mainly for air or water was done \cite{jaluria_gebhart_1974,joshi_gebhart_1987,YuPhysRevE2007,NgJFM2015,ng_ooi_lohse_chung_2017,ng_ooi_lohse_chung_2018,ng_spandan_verzicco_lohse_2020,wangJFM2021,howland_ng_verzicco_lohse_2022}.





According to theoretical work on VC \cite{ShishkinaPhysRevE2016}, there is a boundary point of Prandtl number $Pr=10^{-1}$, where dependencies of convective heat and momentum transport have a transition between the regimes.
Liquids with such low Prandtl numbers are mainly liquid metals.
However, there is not much literature on the low Prandtl number VC.
The work devoted to the study of natural convection in liquid gallium for the crystal growth applications can be distinguished as one of the early ones \cite {braunsfurth_skeldon_juel_mullin_riley_1997}.
In some studies on inclined convection, the position of a cylindrical container at the extreme horizontal point will also correspond to the case of vertical convection \cite{TeimurazovPRF2017,ZwirnerJFM2020}
The most complete study from the fluid mechanics point of view on vertical convection of liquid metal in a box-shaped container was published in \cite {ZwirnerJFM2022}.

Usually convection system depends on the following three non-dimensional input parameters: Rayleigh, Prandtl numbers and aspect ratio: 

\begin{equation}
Ra \equiv \frac{\beta g (T_+ - T_-) L^3}{\nu \kappa},~~~~~~Pr= \frac{\nu}{\kappa},~~~~~~A=\frac{L}{H},
\end{equation}
where $\beta$ is the thermal expansion coefficient, $g$ is the gravity vector, $T_+$ and $T_-$ are the temperature of heated and cooled walls respectively, $L$ and $H$ are the characteristic sizes of container, $\nu$ is the kinematic viscosity and $\kappa$ the thermal diffusivity.


One of the industrial applications associated with the convection of liquid metals is the solidification of pure metals or alloys, which occurs in foundries or casting technologies.
In such technological processes as solidification, temperature and concentration gradients generate thermal or solutal driven convective flows. These convection conditions have a significant and sometimes decisive influence on the formation of the as-cast structure \cite{rappaz2009solidification,WuIJHMT2021}.
To investigate the influence of convection on casting defects (concentration inhomogeneity) formation of a two component alloy, Hebditch and Hunt \cite{Hebditch1974} developed the experimental setup.
This setup consists of a cuboid vessel cooled from one side, while the remaining walls are thermally insulated. 
In such conditions, one vortex circulation similar to VC is formed.

For more precise and complete investigation of such low-Prandtl-number vertical convection related to the solidification process, two recent works were done \cite {BottonIJTS2013, HamzaouiIJTS2019}.
In these works, a box-shaped container with aspect ratios corresponding to the directional solidification benchmark AFRODITE \cite {HachaniIJHMT2012} was chosen.
It should be recognized that such experiment formulation of thermal convection in a liquid metal is very interesting and relevant for the solidification community. However, in these works, the emphasis was placed on studying the possibility of simplifying numerical models to two-dimensional and not on the process of vertical convection itself.

In contrast to the Rayleigh number, widely used in the fluid mechanics community, in works dedicated to convection during solidification, the Grashof number was used as the defining non-dimensional parameter. 
The Grashof number determines the ratio of the buoyancy to viscous force acting on a fluid defined as follows.

\begin{equation}
Gr= \frac{g \beta (T_+ - T_-) H^3}{\nu^2},
\label{Gr_T}
\end{equation}

However, for a more convenient perception of the results by the convection community in appendix \ref{appndx}, the Rayleigh number  is also given for all considered cases.


Thus, by controlling VC, it is possible to influence the casting process and the formation of as-cast structure, which is vital for solidification science.
One of the ways to control convective flows in the liquid phase bulk is electromagnetic (EM) force action. 
This method uses the external alternating magnetic fields to contactlessly generate the EM forces into the liquid metal \cite{EckertEPJST2013}.
To derive the value of electromagnetic forces to a dimensionless form and compare them with buoyancy forces, the dimensionless electromagnetic forcing parameter is introduced \cite {ShvydkiyMMTB2021}.
\begin{equation}
F = \frac{\left< F_{x} \right> H^3}{\rho \nu^2},
\end{equation}
where $\left< F_{x} \right>$ is volume averaged horizontal component of electromagnetic body force and $\rho$ is the mass density.

The balance of buoyancy and electromagnetic forces in such forced convection case is determined using the buoyancy number:

\begin{equation}
N=\frac{F}{Gr} =\frac{\left< F_{x} \right>  }{ \rho g\beta\left(T_{+} - T_{-}\right) },
\end{equation}

Thus, the parameter $N$ 
represents the ratio of two convective forces.
The value of this parameter determines the dominant flow generation mechanism. 
In works \cite {Hachani2015, Avnaim2018a, SariIJTS2021} it is noted that at the $N<1$ corresponds to the case of thermal convection, and $N>1$ is depicted for the case of electromagnetic excitation of the fluid flow.
However, there may be modes with $N$ close to 1, for which there are no exhaustive studies in the literature to date.
Moreover, there are no numerical or experimental studies in which this transition is determined.


Therefore, in this work, we will concentrate only on low Prandtl number cases applied to the directional solidification process.
The purpose of this work is to study the interaction of natural convection and electromagnetically forced flow to determine the transition points in circulating liquid metal flow.




\section{\label{methods}Methods}

The multiphysic model is implemented using a combination of open-source codes of Elmer and OpenFOAM by means of EOF Library coupler  \cite{Vencels2017,Vencels2019}.

\subsection{The governing equations}

\subsubsection{Fluid dynamics equations}

The incompressible Navier–Stokes equations in Oberbeck–Boussinesq approximation
\begin{equation}
\nabla\cdot \mathbf{U}=0
\end{equation}
\begin{equation}
\frac{\partial \mathbf{U}}{\partial t} +\nabla \cdot p + (\mathbf{U} \cdot \nabla) \mathbf{U} - \nu \nabla^2 \mathbf{U}= \frac{\mathbf{F_{b}}+\mathbf{F_{em}}}{\rho_0}
\label{eq:momentum}
\end{equation}
\begin{equation}
\dfrac{\partial (\rho h)}{\partial t} + \nabla \cdot (\rho \mathbf{U} h) + \dfrac{\partial (\rho K)}{\partial t} 
+ \nabla (\rho \mathbf{U} K) - \dfrac{\partial p}{\partial t} = 
\nabla \cdot (\alpha_{eff} \nabla h) + \rho \mathbf{U} \cdot \mathbf{g}
\end{equation}
are solved using finite volume code written in OpenFOAM software by means of the Large Eddy Simulation approach. Here, $\mathbf{U}$ is the velocity vector, $\nu$ is the kinematic viscosity, $\rho_0$ is the mass reference density, $\mathbf{g}$ is the gravity vector, $\mathbf{F}_b$ and $\mathbf{F_{em}}$ depict the buoyancy and electromagnetic forces, $h = e + p/ \rho$ is the enthalpy derived by internal energy $e$, kinetic energy $\mathbf{K} = 0.5 \mathbf{U}^2$ and $\alpha_{eff} = \rho \nu_{t}/Pr_t + k/C_p$ expressed by turbulence kinematic viscosity $\nu_t$, Prandtl number for turbulence mode $Pr_t$, thermal diffusivity $\kappa$ and heat capacity $C_p$   It should be noted, that the temperature field is calculated from these solution variables $e$ and $h$. 

\subsubsection{Electromagnetic field equations}

Electromagnetic forces were calculated using Elmer open-source software. The general equations to calculate the electromagnetic field variable can be presented by
electric scalar $\varphi$ and vector magnetic scalars $\mathbf{A}$ in transient form
\begin{equation}
\Delta \mathbf{A}- \mu \sigma \dfrac{\partial \mathbf{A}}{\partial t} -\mu \sigma \nabla \varphi+\mu \sigma (\mathbf{U}\times \nabla \times \mathbf{A})=-\mu \mathbf{J},	
\label{eq:Aphi}
\end{equation}
\begin{equation}
\mathbf{B} = \nabla \times \mathbf{A},
\label{eq:B}
\end{equation}
\begin{equation}
\mathbf{J} = \sigma \left(-\nabla \varphi + \dfrac{\partial \mathbf{A}}{\partial t} \right),
\label{eq:J}
\end{equation}
and then rewritten in harmonic form using Laplace transformation 
\begin{equation}
\Delta \mathbf{\underline{A}}-j\omega \mu \sigma \mathbf{\underline{A}}-\mu \sigma \nabla \underline{\varphi}+\mu \sigma (\mathbf{U}\times \nabla \times \mathbf{\underline{A}})=-\mu \mathbf{\underline{J}},
\label{eq:Aphi}
\end{equation}
\begin{equation}
\mathbf{\underline{B}} = \nabla \times \mathbf{\underline{A}},
\label{eq:B}
\end{equation}
\begin{equation}
\mathbf{\underline{J}} = \sigma (-\nabla \underline{\varphi} + j \omega \mathbf{\underline{A}}).
\label{eq:J}
\end{equation}
Here, $\omega$ is the angular velocity of the magnetic field expressed as $ 2 \pi f $ using frequency of magnetic field $ f $, $ \mu $ is the absolute magnetic permeability, $ \sigma $ is the conductivity, $\mathbf{J} $ is the current density, $ \mathbf{B}$ is the magnetic flux density $ j$ is the imaginary unit. 
Underlining of a number represents that the number belongs to a complex numbers area. This approach makes it possible to significantly reduce computing resources, because it is not required to calculate magnetic field parameters for each time step.  Magnetic field advection by melt velocity is not taken into account due to a low magnetic Reynolds number ($Re_m=\sigma \mu \langle |u_x| \rangle L < 3.5\cdot10^{-3}$).

\subsubsection{Buoyancy force}

The main driving force in these circumstances is temperature difference and consequently non-equal density in the liquid volume. The buoyancy force depicts $\mathbf{F_b}$ in eq.~\ref{eq:momentum} and is calculated following

\begin{equation}
 \mathbf{F_b}=\mathbf{g} \rho = \mathbf{g} \rho_0[1-\beta(T-T_0)]),
 \end{equation}
 where $\mathbf{g} $ is a gravity vector, $\rho_0$ is the reference density at reference temperature $T_0$.
 
\subsubsection{Electromagnetic force}
The electromagnetic force term in the equation \ref{eq:momentum} is calculated by  
\begin{equation}
\mathbf {F_{em}} =0.5 \Re\{\underline{\mathbf{J}} \times  \underline{\mathbf{B}} \} \end{equation}
at the initial time step of the multi-physic problem. This calculated EM force field is interpolated for finite element mesh in Elmer and then sent to OpenFOAM for the calculating fluid dynamic part of the problem. 







\subsubsection{Boundary conditions and Thermal and kinetic boundary layers (BLs)}

As the thermal boundary conditions on the right and left lateral walls the fixed temperatures $T_-$ and $T_+$ are set.
All remaining walls are thermally insulated. 
Velocity on all walls is set to 0 and the walls are considered as super electrical insulated.

Convection in liquids with small Prandtl numbers is provided by large ratio of the thickness of the thermal to kinetic boundary layers (BLs) (thick thermal BLs and thin kinetic
BLs) \cite{SchumacherPhysRevFluids2016,TeimurazovPRF2017,PhysRevE.82.026318}

The thickness of the thermal boundary layer is defined as follows \cite{SchumacherPNAS2015}:
\begin{equation}
\delta_T = \frac{L}{2 Nu}
\label{delta_T}
\end{equation}

For our cases $50.6<\delta_T<19.9$ mm. Used mesh has a $0.8-0.67$ mm length in horizontal direction, this means the number of nodes  within the thermal BL is always $N_{\delta_T}\geqslant 30$.

The kinetic boundary layer definition from \cite{RevModPhys.81.503,NgJFM2015}:
\begin{equation}
\delta_u = \frac{L}{Re^{1/2}}=\frac{100~mm}{1110^{1/2}}=\frac{100~mm}{33.3}=3 ~mm
\label{delta_u(Re)}
\end{equation}
Thus for our cases number of nodes  within the viscous BL is always $N_{\delta_u}\geqslant 4$.

But if we take wind velocity $ U_{wind} \equiv max \big\{ \langle \overline{u_y}\rangle_{A_{xy}}\big\} $ as a determination parameter then the Reynolds ($7690 \geqslant Re\frac{\langle u_y\rangle_{max} L}{\nu} \geqslant 684$) and BL thickness consequently will lie in the range of $1.14<\delta_u<12~mm$. 
For the cases with a largest $Gr$ or $F$, the number of control volumes was increased up to $150\times90 \times25$.
Therefore, the number of nodes  within the kinetic BL is always $N_{\delta_u}\geqslant 2$.






\subsection{Numerical experiment formulation and considered cases}

In this paper, we have deviated from the classical cubic or cylindrical containers with a unit ratio used in fluid mechanics.
Since the  target area of this work results application is the field of solidification, we turned to the Hebditch and Hunt experiment \cite {Hebditch1974}, which is a classic case for studying the effect of free convection on as-cast structure.
The geometrical aspect ratios of the considered container $A_L= L / H = 5/3 $ and $A_W= W / L = 1/6 $,  where L, W and H are the length, width and height of the cell, respectively. 
Thus, it is a rectangular narrow cell.

\begin{figure}[h]
\includegraphics[width=1.0\linewidth]{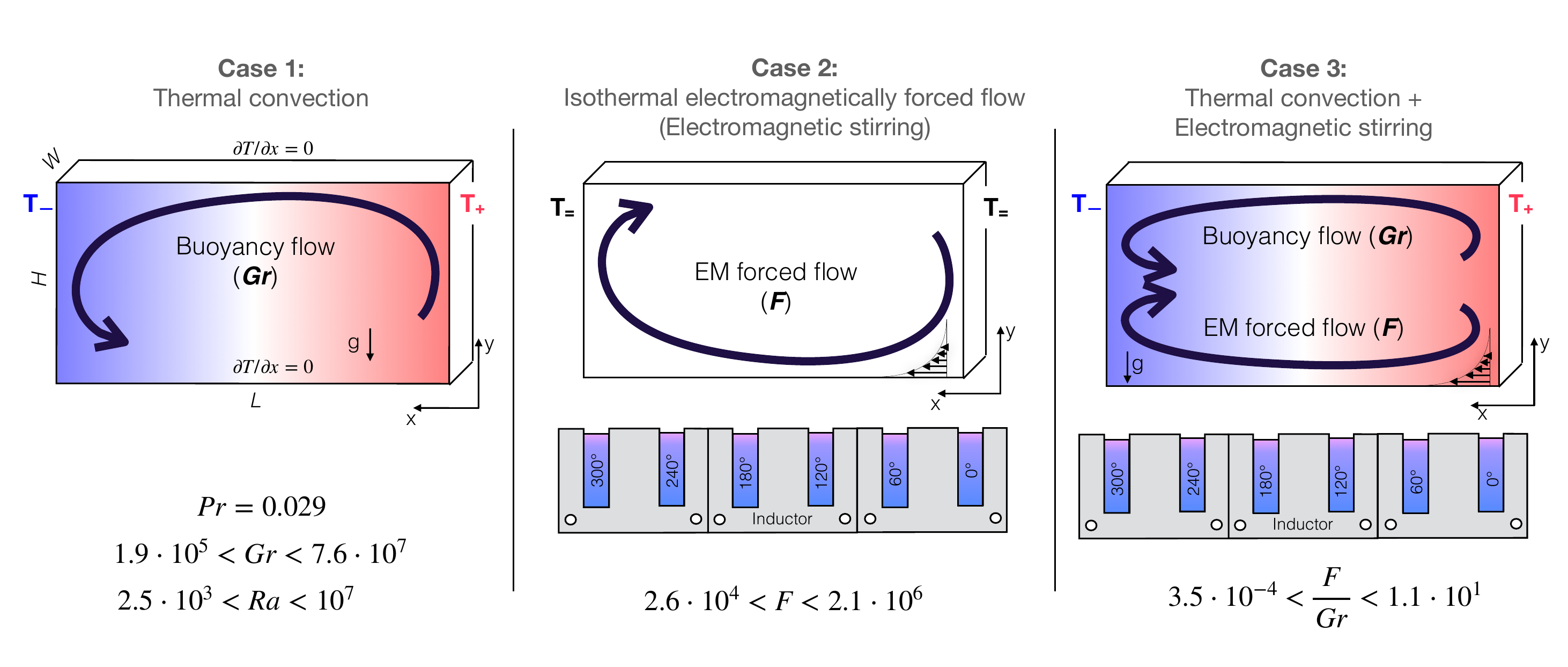}
\caption{\label{cases} Considered cases sketches.}
\end{figure}

To create a horizontal temperature gradient, the left side wall is cooled ($T_-$) and the opposite wall is heated up ($T_+$). These circumstances are a pure VC and correspond to the Case 1 (Fig. \ref{cases}).
For the second case $T_-=T_+$, so we have an isothermal liquid. 
Additionally, the magnetic field inductor is placed under the cell. 
This inductor generates EM body forces into the liquid metal, under the influence of which electromagnetic stirring will occur.
And in the third case a combination of both buoyancy and EM forces is investigated.

\subsection{Experimental validation}

\begin{figure}[h]
\includegraphics[width=0.4\linewidth]{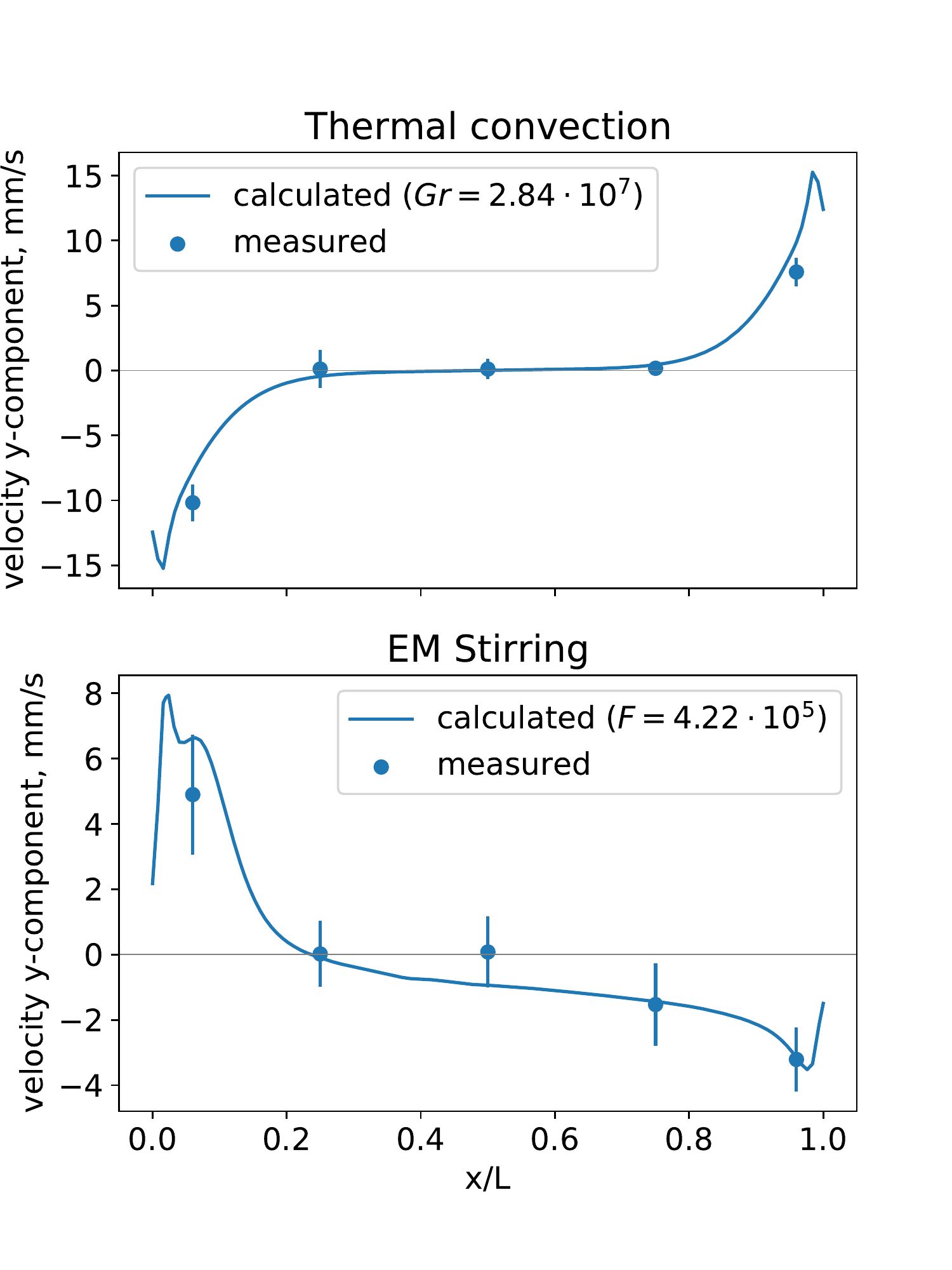}
\caption{\label{UDVvalidation} Vertical thermal convection and electromagnetic stirring simulation validation by UDV measurements liquid metal velocities.}
\end{figure}

To validate created models, an experimental setup was built up.
This experimental setup allows to the creation a horizontal thermal gradient and EM forces. GaInSn is chosen as a working liquid and properties of which are taken according to \cite{Plevachuk2014,MusaevaStPPUJ2016,ZuernerJFM2020,YangPhysRevFluids2021}. 
The vertical component of liquid metal flow velocity was measured by the 5 Ultrasound Doppler Velocimetry probes.
Comparison of measured and calculated velocity profiles for the thermal convection and EM stirring is shown in Fig. \ref{UDVvalidation}.
As can be seen from the graphs, the numerical model quite accurately predicts both the shape of the profiles and the velocity magnitude.

\section{\label{results}Results}

\subsubsection{Flow in the vertical convection case (Case 1)}

\begin{figure}[h]
\includegraphics[width=0.5\linewidth]{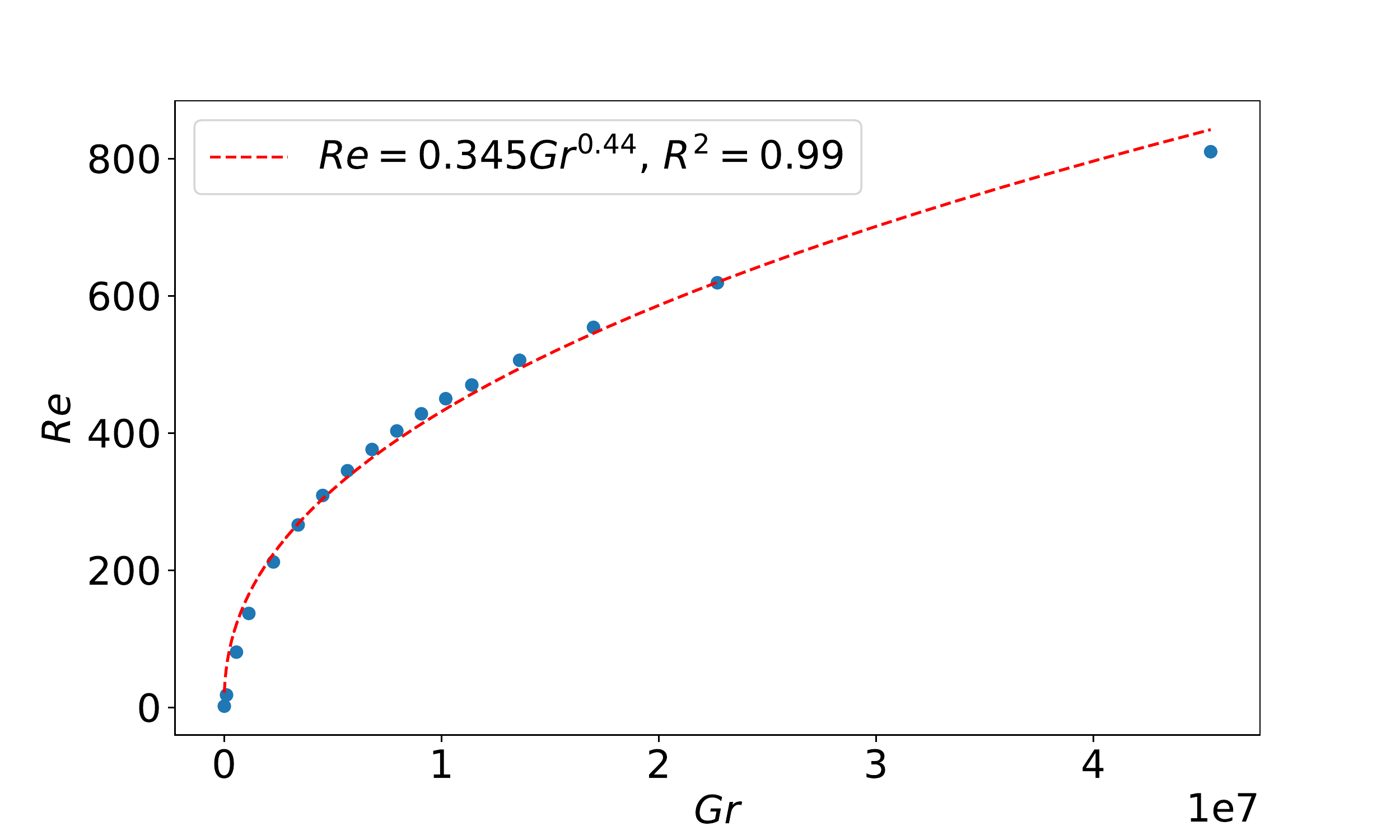}
\caption{\label{Re(Gr)aprx} Dependence of the Reynolds number on the Grashof number.}
\end{figure}
The first considered case is a thermal vertical convection.
A parametric study of the effect of the Grashof number on the liquid metal flow was carried out.
In this case, the Grashof number varied in the range $ 1.89 \cdot 10^5 <Gr <7.57 \cdot 10^7 $.
If we plot the obtained values of the Reynolds number over the Grashof number, we get the dependence shown in Fig.~\ref{Re(Gr)aprx}.
It can be seen that with an increase in the Grashof number, the Reynolds number also grows as expected.
Moreover, for small values of the Grashof number, the Reynolds number increases rapidly and after $ Gr = 0.5 e ^7 $ the growth slows down.
The data obtained is approximated by the power function $Re=0.345 Gr^{0.44}$ with a sufficiently high accuracy ($R^2=0.99$).

\subsubsection{Electromagnetically driven isothermal flow (Case 2)}

\begin{figure}[h]
\includegraphics[width=0.5\linewidth]{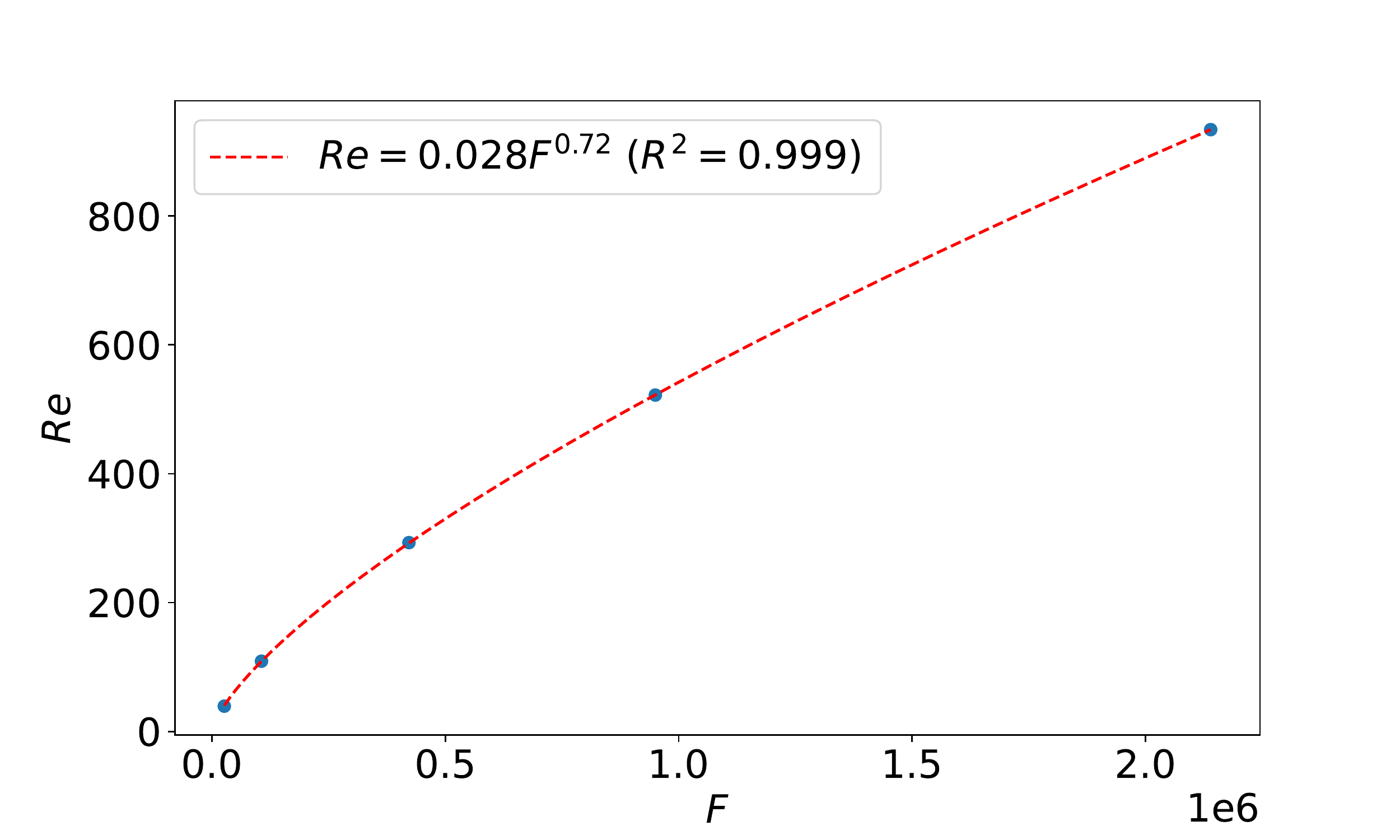}
\caption{\label{Re(F)aprx} Dependence of the Reynolds number on the Electromagnetic forcing parameter.}
\end{figure}

On Fig.~\ref{Re(F)aprx} the dependency of the Reynolds number on the EM forcing parameter is shown. Only 5 cases were calculated, but this was enough to clearly see the dependence. The calculated values can be described by the power function $ Re (F) = 0.028 F^{0.72} $ with the coefficient of determination $ R^2 = 0.999 $.
A similar dependence was obtained in the work of Avnaim et al \cite {Avnaim2018} ($Re(F)=0.31F^{0.5}$).
The difference between the two approximating functions can be explained by different metals (pure Ga in work \cite {Avnaim2018}) and cavity widths.

\subsubsection{Combined buoyancy and EM driven flow (Case 3)}

The two cases considered above have one acting body force. In this work what it is more interesting are the results of calculating the liquid metal flow under the simultaneous influence of buoyancy and electromagnetic forces.

\begin{figure*}
\includegraphics[width=1\linewidth]{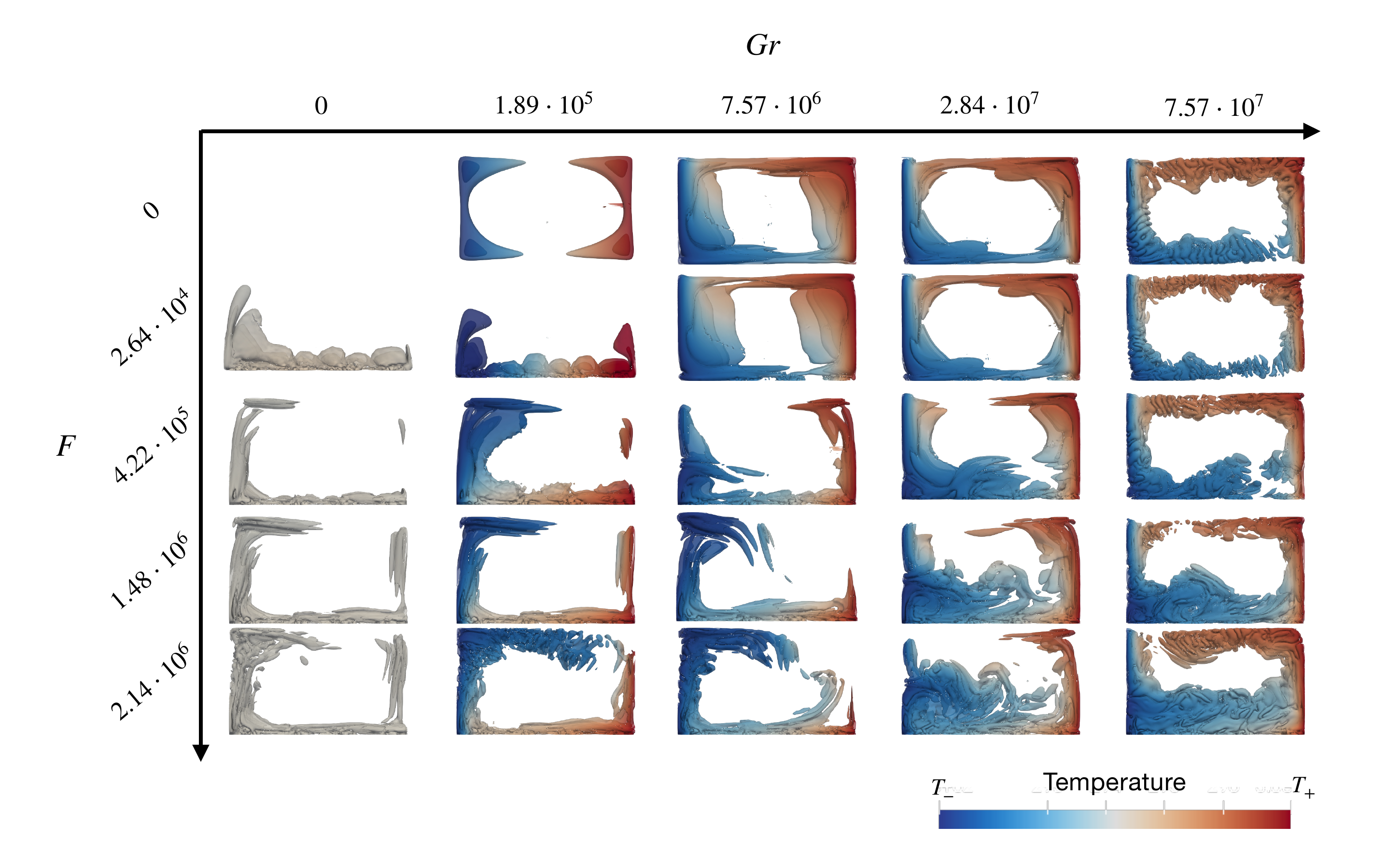}
\caption{\label{Gr_F_figure} Instantaneous flow patterns at different Grashof number ($Gr$) and EM forcing parameter ($F$).}
\end{figure*}
Fig.~\ref {Gr_F_figure} shows a map of the flow patterns for various combinations of the electromagnetic forcing parameter ($F$) and the Grashof number ($Gr$).
These flow patterns are plotted as a Q-criterion.
The color indicates the temperature of the liquid.
The upper horizontal line corresponds to Case 1, i.e. in this case the electromagnetic forces are equal to zero ($ F = 0 $). The left wall cools down and the right one heats up, forming a horizontal temperature gradient.
The cooled volumes of liquid descend along the left wall and are transported towards the opposite wall.
At the same time, the heated, less dense metal rises under the influence of buoyancy forces along the right wall.
In such circumstances, a counterclockwise single-vortex circulation is formed.
Also, with an increase in the Grashof number, the temperature advection increases.
When cold or warm volumes of liquid are transported with increasing velocity, convection heat transfer will predominate over conduction.

Case 2 corresponds to the left column of the map shown in Fig.~\ref {Gr_F_figure}.
Since in this case all the boundary conditions are set as thermal insulation, the temperature throughout the volume will be the same.
In contrast to the VC case, the flow pattern during electromagnetic stirring is slightly asymmetrical.
This is due to the fact that electromagnetic forces are applied only at one wall and not at two.
Another thing that is worth noting for the first two cases is that with an increase in EM forcing parameter and the Grashof number, as follows from Fig.~\ref{Re(Gr)aprx} and \ref{Re(F)aprx}, the value of the Reynolds number increases and we can observe the formation of small-scale vortices at $ F = 2.14 \cdot 10 ^ 6 $ or $ Gr = 7.57 \cdot 10 ^ 7 $. The flow pattern in the whole range of $Gr$ or $F$ remains the same.

However, if we combine electromagnetic stirring and vertical convection, we can affect the flow pattern.
To determine the interaction of VC with EM stirring, we solved all combinations of case 1 with case 2.
As a result, the matrix or map for the mutual action of electromagnetic forces ($ F $) and buoyancy forces ($ Gr $) is shown in Fig.~\ref {Gr_F_figure} below the first line and to the right of the left column.
With the minimum value of the electromagnetic forcing parameter ($ F = 2.64 \cdot 10^4 $), a significant effect on the flow pattern was exerted only in for $ Gr = 1.89 \cdot 10^5 $. Whereas in other cases, the magnitude of the electromagnetic forces is not enough to affect the process of vertical convection.
However, if the forces are increased to $ F = 4.22 \cdot 10^5 $, then in most cases the result of the influence of EM forces on the flow pattern is visible.
Thus, electromagnetic stirring will dominate with the Grashof number $ Gr = 1.89 \cdot 10^5 $ at $ F \leqslant 4.22 \cdot 10^5 $.

In cases where developed convection and insignificant EM forces ($F=4.22\cdot10^5$ and $Gr=2.84\cdot10^7 $; $F=4.22\cdot10^5$ and $Gr=7.57\cdot10^7 $; $F=1.48\cdot10^6 $ and $Gr=7.57\cdot10^7 $) you can see how these forces form an undeveloped vortex at the bottom surface. In such a regime, this is an obstacle for the main vortex and, presumably, braking of the convection flow may occur.
Not so pronounced, due to the lower buoyancy force density than the electromagnetic, reverse mode of braking by convection of electromagnetic stirring in the cases $F=1.48\cdot10^6$ and $Gr=7.57\cdot10^7$; $F=2.14\cdot10^6$ and $Gr=7.57\cdot10^7$ can be seen. In this mode, the vortex opposing the main vortex is formed in the upper right region of the cell where the EM-driven flow velocity is less.

If we look at the results located along the diagonal as $F$ and $Gr$ increase, we can see that a pronounced two-vortex flow structure is formed.
The lower vortex is caused by electromagnetic forces and is directed clockwise, while the upper vortex is caused by buoyancy forces and is directed in the opposite direction.
Since electromagnetic stirring and vertical convection simultaneously coexist in this regime, it can be assumed that such a $F/Gr$ ratio is the boundary. But it is necessary to analyze this detected border in more detail.

\subsubsection{Dependence of Reynolds and Nusselt numbers on the buoyancy number $N=F/Gr$}

For a more detailed analysis of the combined effect of electromagnetic and buoyancy forces, we turned to the two main integral characteristics of convection, the Reynolds and Nusselt numbers. These parameters  describe the intensity of convective flow and the effective heat transport \cite{TeimurazovPRF2017}.

Since our cell is longer in x direction and, moreover, thermal gradient is applied along x axis, we will evaluate the \textbf{Reynolds number} in the following way:
\begin{equation}
Re=\frac{\langle |v_x|\rangle H}{\nu}
\label{Re_x_avg}
\end{equation}
where $u_x$ is the velocity horizontal component and $\langle  T\rangle_x$ denotes averaging over any plane $x = const$ and over time. 


\begin{figure}[h]
\includegraphics[width=0.5\linewidth]{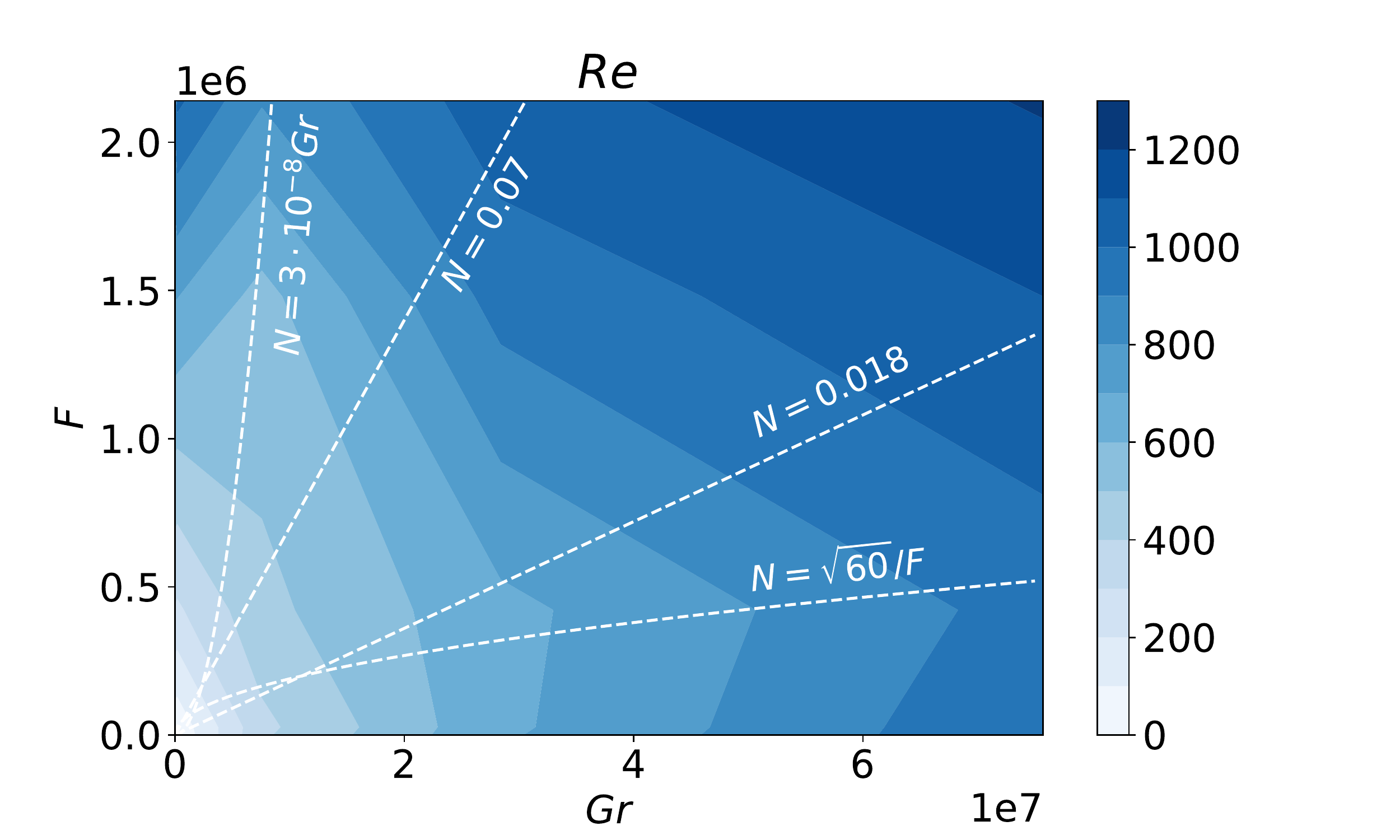}
\caption{\label{Re(N)count} Dependence of the Reynolds number on the EM forcing parameter and the Grashof number.}
\end{figure}

The contour plot of the Reynolds number over the $F$ and $Gr$ numbers is shown on Fig. \ref{Re(N)count}.
As expected, as $F$ and $Gr$ increase, the Reynolds number grows as well.
However, it does not increase uniformly.
It can be noted that the growth of the Renolds number slows down in two regions.
These regions can be roughly approximated by power functions $N (F) = 60^{\frac{1}{2}}/F$ and $ N (Gr) = 3 \cdot 10^{-8} Gr$.
These areas correspond to the braking modes noted earlier. Indeed, with such ratios $ F/Gr $, EM driven flow will slow down buoyancy driven and vice versa.
It should also be noted that in addition to braking, there is an area of a slight increase in the Reynolds number.
It lies in the range $ 0.018 <F / Gr <0.07 $ and corresponds to the regime when there are 2 developed vortices that do not slow down, but rather slightly increase the velocity of each other.

\subsubsection{Nusselt number}

Convective heat transfer is analyzed by means of the \textbf{Nusselt number}:

 \begin{equation}
Nu=\frac{\langle u_x T\rangle_x - \kappa \partial_x \langle  T\rangle_x }{\chi (T_+-T_-) L^{-1}}
\label{Re_max}
\end{equation}
with, $\kappa$ -- thermal diffusivity of the fluid \cite{PhysRevLett_Shishkina,TeimurazovPRF2017}.

\begin{figure}[h]
\includegraphics[width=0.5\linewidth]{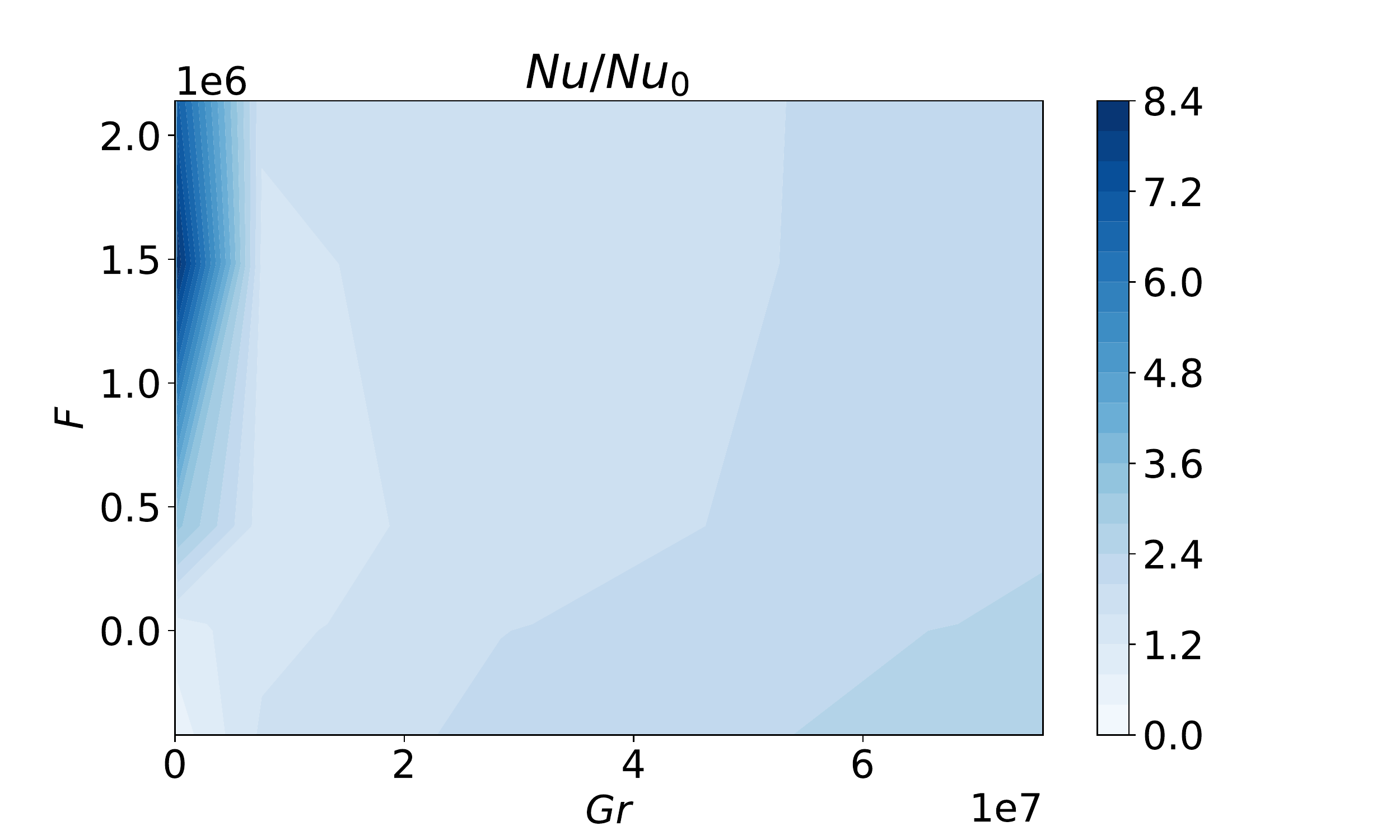}
\caption{\label{Nu(N)count} Dependence of the relative Nusselt number on the EM forcing parameter and the Grashof number (Global heat and momentum transport).}
\end{figure}

Fig.~\ref{Nu(N)count} shows the contour plot of Nusselt number in relation to cases without electromagnetic forcing ($F=0$).
Almost in the entire investigated area, there are no significant changes in this parameter.
As the Grashof number increases, the Nusselt number gradiently increases, reaching values greater than 2.
This analysis also considers the $F<0$ zone and in this area the Nusselt number increases, since the electromagnetic flow additionally enhances convection.
But the most pronounced effect of electromagnetic force on convective heat transport is observed at low values of the Grashof number ($Gr<1\cdot10^7$) and high values of electromagnetic forces ($F>5\cdot10^5$).
In this region, electromagnetic stirring dominates, and even at low values of $ Gr $, good convective heat transfer is ensured.
Moreover, this region is described by the curve $ N(Gr)=3\cdot 10^{-8}Gr$ shown in Fig. \ref {Re(N)count}.
For the VC, in work \cite {ShishkinaPhysRevE2016}, the dependence $Nu = Pr^{0.25} Pr^{0.25}$ is derived. In the cases $ Gr> 2 \cdot10^6 $, the deviation of our results from theoretical ones was less than 15 \%.

\begin{figure}[h]
\includegraphics[width=0.5\linewidth]{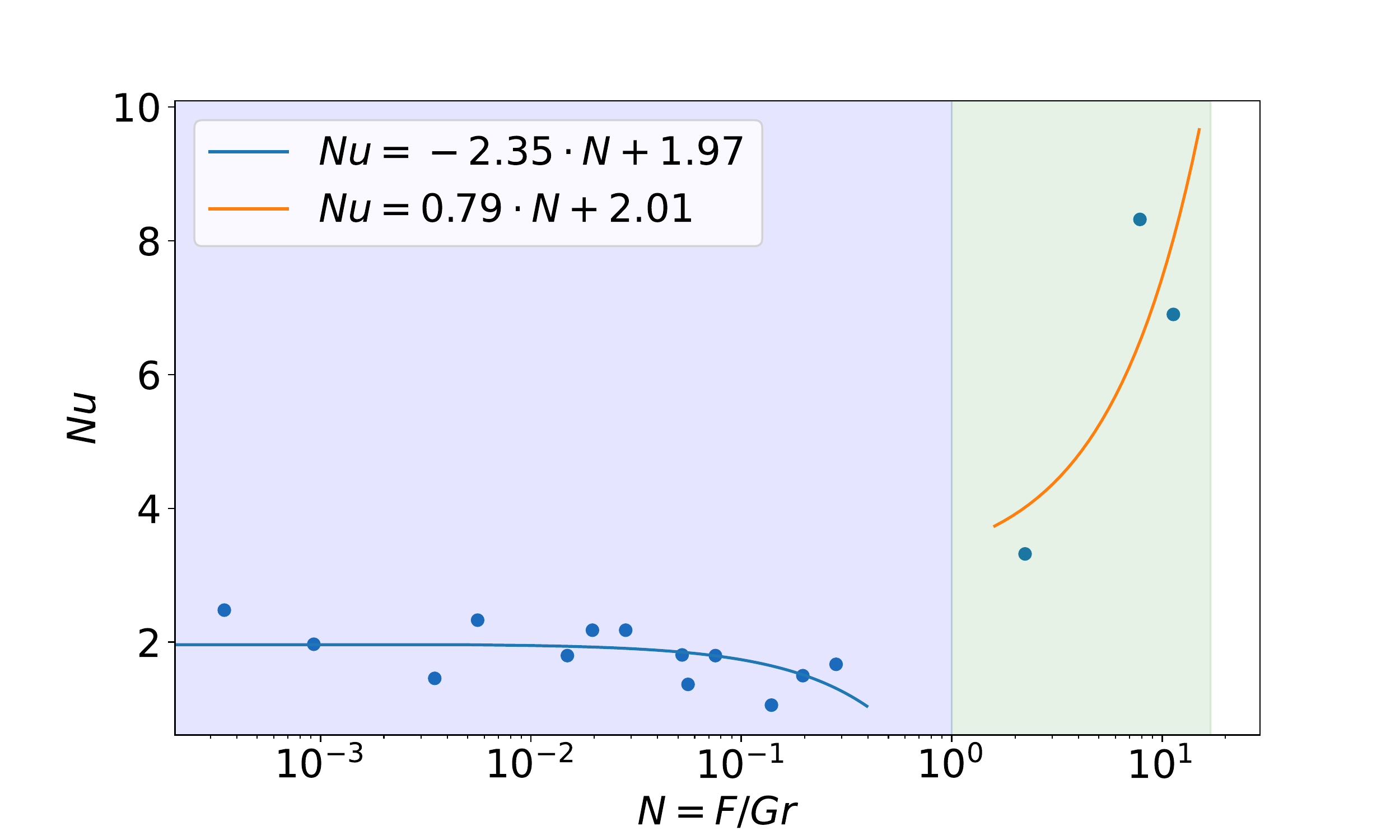}
\caption{\label{Nu(N)} Dependence of the  Nusselt number on the $N$ parameter.}
\end{figure}

A more informative picture of the mutual influence of $F$ and $Gr$ on the Nusselt number can be obtained from the Fig.~\ref {Nu(N)}.
Scatter plot of Nusselt number over the $N = F/Gr$ ratio is shown on this graph.
It is clearly observed that the graph can be divided into 2 parts.
In the first part at the $ N <1 $, the values of the Nusselt number do not exceed three.
An approximation of the data obtained in this region ($ Nu = -2.35 \cdot N + 1.97 $) shows that with a decrease in $ N $ the Nusselt number slightly increases.
This can be explained by the fact that by increasing the electromagnetic forces, convection velocity decreases.
However, when the electromagnetic forces exceed the buoyancy forces ($F>Gr$; $N>1$) the Nusselt number increases several times and reaches $\sim8$. Although there are only 3 data cases in this area, they can be approximated by the function $ Nu = 0.79\cdot N + 2.01 $

\subsubsection{Isotherms deformation analysis}

The last qualitative parameter for analysis is the degree of deformation of the isotherms.
This parameter comes from one of the applications of such a convection system -- solidification.
In vertical convection, it is typical that the flow of liquid metal transfers temperature and deforms the isotherms.
For example, in the middle of the length of the cell in the upper area, the temperature will be higher than in the lower one.
During the phase transition, the temperature value is important.
Under the natural convection circumstances due to this temperature distribution unequal solidification conditions will be generated in the lower and upper regions.
In some cases, it is useful to select such a mode in which the deformation of the isotherms will be minimal \cite {Avnaim2018a}.

\begin{figure*}
\includegraphics[width=14cm]{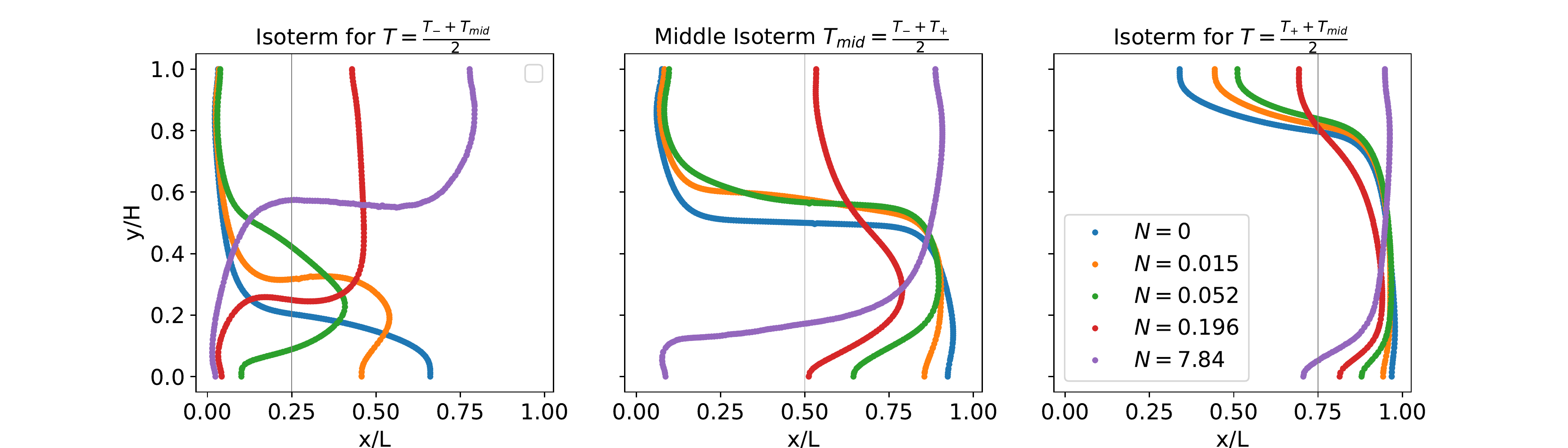}
\includegraphics[width=14cm]{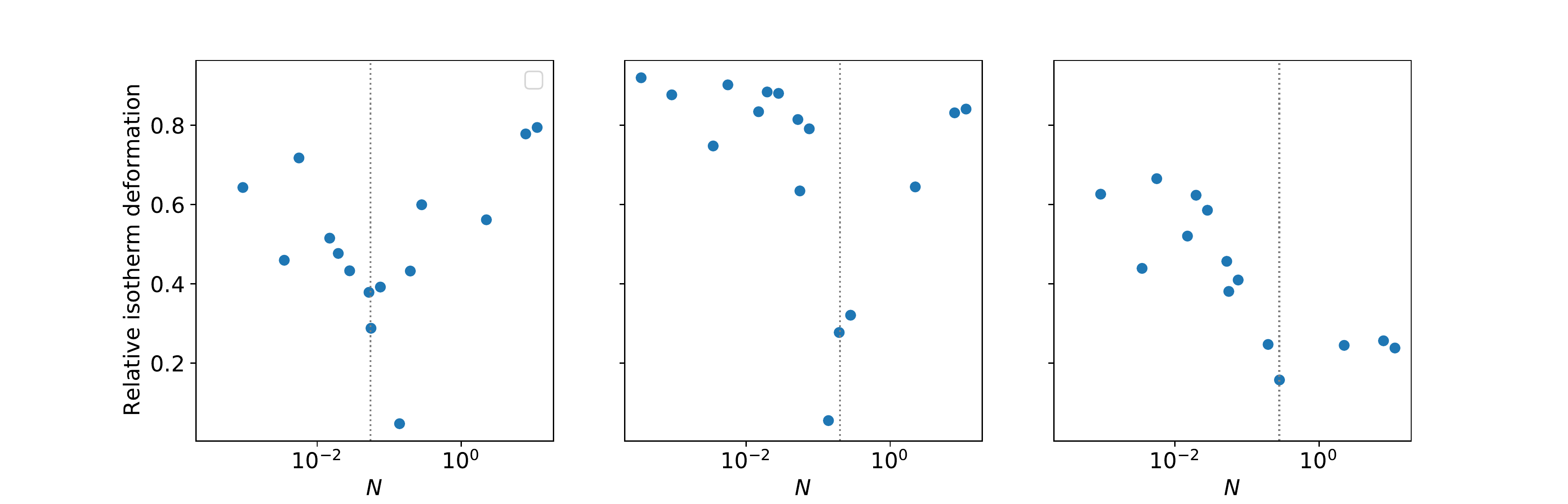}
\caption{\label{isotherms}Isotherms deformation for different forcing conditions.}
\end{figure*}

For such an assessment, we took three isotherms: the middle one, the one closest to the cooled wall, and the one closest to the heated wall.
Time-averaged isotherm curves for selected cases are shown in Fig.~\ref {isotherms} (top line).
The middle picture corresponds to the average temperature isotherm, the left isotherm located closer to the cooled wall and the right isotherm located closer to the heated wall.
In all three cases, a symmetric isotherm can be observed for the case without electromagnetic forcing ($ N = 0 $).
As the electromagnetic forces increase, the position of this line shifts in the direction against natural convection.
In the case of $ N = 0.015 $, the electromagnetic forces are not enough to make significant changes in the shapes of isotherms, only slightly deforming them in the lower region.
For the cases at $ N = 0.052 $ and $ N = 0.196 $, a pronounced deformation of isotherms by EM-driven flow is already observed. For the middle isotherm, bending of the curves in the central region is visible, which indicates a two-vortex flow pattern.
At $ N = 7.84 $, the complete dominance of electromagnetic mixing is observed; in this case, the positions of the isotherms are opposite to the case of $ N = 0 $.

The projection of these isotherms on the x-axis is the deformation of the isotherm. The dimensionless of this parameter is carried out by dividing by the cell length (L).
The bottom row of plots in Fig.~\ref {isotherms} demonstrates the dependence of this dimensionless parameter of deformation of isotherms over the $ N $ parameter.
In all three cases, a negative peak is observed, which indicates the minimum deviation of the isotherm shape from a straight line.
This means that the balance of EM driven and natural convection flows at this $N$ value is close to equal.
For an isotherm shifted towards the cooled wall this minimum peak exists at $ N \sim 0.05 $. In the case of the middle isotherm, this peak is observed at $ N \sim0.2 $. And in the last considered isotherm case, the peak placed at $ N = 0.3 $.
It can be seen how an increase in the temperature of the isotherm pushes this peak towards higher values of the parameter $ N $.
This is due to the fact that electromagnetic forces act on the metal only from the other side, forming a non-symmetric flow.




\section{Conclusion}

\begin{itemize}
    \item Flow patterns for different F/Gr ratios were obtained.
 \item The analysis of the Reynolds number showed that under the action of electromagnetic forces on natural convection, there are two modes of braking of the liquid metal flow and a region where the Reynolds number increases.
  This region lies in the range $ 0.018 <N <0.07 $ and is a border zone in the transition from the dominance of convection over electromagnetic stirring and vice versa.

 \item Analysis of convective heat transfer has shown that at a value of $ 0 <N <1 $ the Nusselt number changes insignificantly. Whereas at values of $ N> 1 $, a sharp increase in the Nusselt number appears.
 Thus, the transition point for the convective heat transfer is located at $ N = 1 $.
 Regime $ N (Gr)> 3 \cdot 10^{-8} Gr $ is characterized by a significant increase in convective heat transfer.

 \item Isotherms deformations are analyzed. 
The effect of the balance of electromagnetic forces on the relative deformation of the isotherm depends on the temperature and lies in the range $ 0.05 <N <0.2 $.
For the case of directional horizontal solidification, this boundary value is close to $ N = 0.05 $.
\item The open source computation code can be found here \footnote{\url{https://github.com/shvydkiy/EM-Stirring-Convection-Solidification/tree/main/EM\%2BThermalConvection}}.
 \item Further works will be focused on the implementation with the solidification of AFRODITE experimental benchmark case.
  \item Investigation of the influence of the density of electromagnetic forces expressed by the magnetic field penetration depth or the shielding parameter is also of interest.

\item All results presented in this work were obtained at a fixed Prandtl number. Since this dimensionless parameter is one of the determinants of convection, it is of certain interest to determine what effect a change in the Prandtl number will make on the results obtained. This will give an idea to which types of liquid metals the results are applicable.
Moreover, based on the results of this work, a proposal is made to investigate the $F/Ra$ ratio instead of $F/Gr$. It can more accurately reflect the nature of the flow in liquid metals, since it takes into account the Prandtl number. But this hypothesis lies outside the scope of this work and needs additional investigation.

\end{itemize}

\begin{acknowledgments}
The present work was completed within the framework of DAAD funding programme: One-Year Grants for Doctoral Candidates, 2020/21 (ID 57507870).
The authors wish to acknowledge Alexander Köppen and Christian Görsch for preparing experimental setup.
\end{acknowledgments}

\bibliography{apssamp}

\newpage
\appendix

\section{Simulation parameters}
\begin{table}[h]
\caption{\label{appndx}
Simulation parameters and measured quantities : Grashof number ($Gr$), electromagnetic forcing parameter ($F$), buoyancy number ($N=F/Gr$), Nusselt number ($Nu$), Rayleigh number ($Ra$). The andtl number is fixed to $Pr = 0.029$.}
\begin{ruledtabular}
\begin{tabular}{ccccccc}
$Gr$ & $F$ & $N$ & $Re$ & $Nu$& control volumes& $Ra_L$ \\
\hline
\\
$1.89 \cdot 10^4$ & 0  & - & $2.26 \cdot 10^0$ & 0.988 & $125\times 75 \times 25$ & $2.54\cdot10^3$ \\
$1.89 \cdot 10^5$ & 0  & - & $2.24 \cdot 10^1$ & 1 & $125\times 75 \times 25$ & $2.54\cdot10^4$ \\
$1.89 \cdot 10^6$ & 0  & - & $1.71 \cdot 10^2$ & 1.15 & $125\times 75 \times 25$ & $2.54\cdot10^5$ \\
$7.57 \cdot 10^6$ & 0  & - & $3.86 \cdot 10^2$ & 1.48 & $125\times 75 \times 25$ & $1.02\cdot10^6$ \\
$1.51 \cdot 10^7$ & 0  & - & $5.32 \cdot 10^2$ & 1.72 & $125\times 75 \times 25$& $2.03\cdot10^6$  \\
$2.84 \cdot 10^7$ & 0  & - & $6.87 \cdot 10^2$ & 1.99 & $125\times 75 \times 25$ & $3.81\cdot10^6$ \\
$7.57 \cdot 10^7$ & 0  & - & $9.93 \cdot 10^2$ & 2.51 & $125\times 75 \times 25$& $1.02\cdot10^7$  \\
\\
0 & $2.64\cdot 10^4$  & - & $3.92 \cdot 10^1$ & - & $125\times 75 \times 25$  &- \\
0 & $1.06\cdot 10^5$  & - & $1.09 \cdot 10^2$ & - & $125\times 75 \times 25$  &- \\
0 & $4.22\cdot 10^5$  & - & $2.93 \cdot 10^2$ & - & $125\times 75 \times 25$ &-  \\
0 & $9.5\cdot 10^5$  & - & $5.22 \cdot 10^2$ & - & $125\times 75 \times 25$ &-  \\
0 & $1.48\cdot 10^6$  & - & $7.06 \cdot 10^2$ & - & $125\times 75 \times 25$  &- \\
0 & $2.14\cdot 10^6$  & - & $1 \cdot 10^3$ & - & $125\times 75 \times 25$ &-  \\
0 & $2.64\cdot 10^6$  & - & $1.26 \cdot 10^3$ & - & $150\times 90 \times 25$ &-  \\

\\
$1.89 \cdot 10^5$ & $2.64\cdot 10^4$  & $1.39\cdot 10^{-1}$ & $3.43 \cdot 10^1$ & 1.06 & $125\times 75 \times 25$ & $2.54\cdot10^4$ \\
$7.57 \cdot 10^6$ & $2.64\cdot10^4$  & $3.49\cdot10^{-3}$  & $3.76\cdot10^2$ & 1.46 & $125\times75\times25$ & $1.02\cdot10^6$ \\
$2.84 \cdot 10^7$ & $2.64\cdot10^4$  & $9.29\cdot10^{-4}$ & $6.8\cdot10^2$ & 1.97 & $125\times75\times25$ & $3.81\cdot10^6$ \\
$7.57 \cdot 10^7$ & $2.64\cdot10^4$  & $3.49\cdot10^{-4}$ & $9.91\cdot10^2$ & 2.48 & $150\times90\times25$ & $1.02\cdot10^7$ \\

$1.89 \cdot 10^5$ & $4.22\cdot 10^5$  & $2.23\cdot 10^{0}$ & $2.85 \cdot 10^2$ & 3.32 & $125\times 75 \times 25$ & $2.54\cdot10^4$ \\
$7.57 \cdot 10^6$ & $4.22\cdot 10^5$  & $5.58\cdot10^{-2}$ & $4.72\cdot10^2$ & 1.37 & $125\times75\times25$ &$1.02\cdot10^6$  \\
$2.84 \cdot 10^7$ & $4.22\cdot 10^5$  & $1.49\cdot10^{-2}$ & $6.74\cdot10^2$ & 1.8 & $125\times75\times25$ & $3.81\cdot10^6$ \\
$7.57 \cdot 10^7$ & $4.22\cdot 10^5$  & $5.58\cdot10^{-3}$ & $9.42\cdot10^2$ & 2.33 & $150\times90\times25$& $1.02\cdot10^7$  \\

$1.89 \cdot 10^5$ & $1.48\cdot 10^6$  & $7.84\cdot 10^{0}$ & $7.03\cdot10^2$ & 8.32 & $125\times 75 \times 25$ & $2.54\cdot10^4$  \\
$7.57 \cdot 10^6$ & $1.48\cdot 10^6$  & $1.96\cdot10^{-1}$ & $5.68\cdot10^2$ & 1.5 & $125\times75\times25$ & $1.02\cdot10^6$\\
$2.84 \cdot 10^7$ & $1.48\cdot 10^6$  & $5.23\cdot10^{-2}$ & $9.41\cdot10^2$ & 1.81 & $125\times75\times25$ & $3.81\cdot10^6$  \\
$7.57 \cdot 10^7$ & $1.48\cdot 10^6$  & $1.96\cdot10^{-2}$ & $1.1\cdot10^3$ & 2.18 & $150\times90\times25$ & $1.02\cdot10^7$  \\
 
$1.89 \cdot 10^5$ & $2.14\cdot 10^6$  & $1.13\cdot 10^{1}$ & $1.02\cdot10^3$ & 6.9 & $150\times90\times25$ & $2.54\cdot10^4$  \\
$7.57 \cdot 10^6$ & $2.14\cdot 10^6$  & $2.82\cdot10^{-1}$ & $8.08\cdot10^2$ & 1.67 & $150\times90\times25$ & $1.02\cdot10^6$  \\
$2.84 \cdot 10^7$ & $2.14\cdot 10^6$  & $7.53\cdot10^{-2}$ & $1.06\cdot10^3$ & 1.8 & $150\times90\times25$ & $3.81\cdot10^6$ \\
$7.57 \cdot 10^7$ & $2.14\cdot 10^6$  & $2.82\cdot10^{-2}$ & $1.21\cdot10^3$ & 2.18 & $150\times90\times25$ & $1.02\cdot10^7$  \\

$1.89 \cdot 10^5$ & $-4.22\cdot 10^5$  & $-2.23\cdot 10^{0}$ & $3.02\cdot10^2$ & 0.58 & $125\times75\times25$ & $2.54\cdot10^4$   \\
$7.57 \cdot 10^6$ & $-4.22\cdot 10^5$  & $-5.58\cdot10^{-2}$ & $5.25\cdot10^2$ & 1.67 & $125\times75\times25$ & $1.02\cdot10^6$  \\
$2.84 \cdot 10^7$ & $-4.22\cdot 10^5$  & $-1.49\cdot10^{-2}$ & $7.72\cdot10^3$ & 2.12 & $150\times90\times25$ & $3.81\cdot10^6$ \\
$7.57 \cdot 10^7$ & $-4.22\cdot 10^5$  & $-5.61\cdot10^{-3}$ & $1.22\cdot10^3$ & 2.64 & $150\times90\times25$ & $1.02\cdot10^7$  \\
\end{tabular}
\end{ruledtabular}
\end{table}

\newpage
\section{Some considered vertical convection cases.}
\begin{table}[h]
\caption{\label{appndxRev}
Some considered vertical convection cases.}
\begin{ruledtabular}
\begin{tabular}{cccccc}
Ref. & $Pr$ & $Ra$ (or $Gr$) & $\Gamma$ & $Nu$ & $Re$ 
\\
\hline
\\

\cite{braunsfurth_skeldon_juel_mullin_riley_1997} &  $Pr=0.02$  & $10^3<Gr<10^5$ & $H/L=1$ & -&- \\ 

\cite{YuPhysRevE2007} & $Pr=0.71$   & $1<Ra<10^8$& $\Gamma=0.5-20$ & $1-10^2$& -\\

\cite{BottonIJTS2013} &  $0.0045<Pr< 0.03$  & $1.3\cdot10^6<Gr<1.6\cdot10^7$&$L/H=10/6$ & - & - 
\\ 

\cite{NgJFM2015} &  $Pr=0.709$ (air)  &$10^5<Ra<10^9$ &$L_x/H=8$; $L_y/H=4$ & - & $\sim 10^2 - 10^4$
\\ 

\cite{ShishkinaPhysRevE2016} & $10^{-2}<Pr<30$ & $ 10^5<Ra<10^{10}$ & L/D= 1 &  $\sim 10^0 - 10^2 $ &
\\

\cite{TeimurazovPRF2017} & $Pr=0.0083$ & $Ra=(4.8-7.7)\cdot 10^6 $ & $L/D=5$ &  $\sim 25 - 35 $ &$\sim6\cdot 10^6$
\\

\cite{ng_ooi_lohse_chung_2017} & $Pr=0.709$ (air)  &$10^5<Ra<10^9$&$L_x/H=8$; $L_y/H=4$ & $0.02 < \frac{Nu}{Ra^{\frac{1}{3}}}<0.1$&$0.2 < \frac{Re}{Ra^{\frac{1}{2}}}<1.8$ \\ 

\cite{HamzaouiIJTS2019} &$10^{-3}<Pr<10$  & $10^3 <Gr< 5 \cdot 10^7$  &$L/H=10/6$  & -&- \\ 

\cite{ZwirnerJFM2020} &$Pr \sim 0.009$  & $Ra>10^7$ & $L/D=1$& $ \sim 6-7$&$ \sim 10^4$  \\

\cite{wangJFM2021} & $Pr=10$ & $10^7<Ra<10^{14}$  & $H/L=1$ & $17-1908$ &$\sim 10^1 - 10^4$
\\

\cite{ZwirnerJFM2022} & $Pr=0.03$ & $5\cdot 10^3<Ra<10^8$ & 1, 2, 3 and 5. &  $\sim 1 - 20 $ &$\sim10^2-2\cdot10^4$
\\
\end{tabular}
\end{ruledtabular}
\end{table}




\nocite{*}

\end{document}